\documentclass[%
 aip,
 amsmath,amssymb,
 reprint,%
]{revtex4-1}

\usepackage{graphicx}
\usepackage{dcolumn}
\usepackage{bm}

\usepackage[utf8]{inputenc}
\usepackage[T1]{fontenc}
\usepackage{mathptmx}
\usepackage{etoolbox}
\usepackage{hyperref}
\hypersetup{
    colorlinks = true,
    linkcolor = blue,
    citecolor = blue,
    urlcolor = blue,
    pdfborder = {0 0 0},
    pdfusetitle = true
}

\makeatletter
\def\@email#1#2{%
 \endgroup
 \patchcmd{\titleblock@produce}
  {\frontmatter@RRAPformat}
  {\frontmatter@RRAPformat{\produce@RRAP{*#1\href{mailto:#2}{#2}}}\frontmatter@RRAPformat}
  {}{}
}%
\makeatother
\begin{document}

\preprint{AIP/123-QED}

\title{
Substitutional platinum as an efficient nonradiative recombination center in silicon
}
\author{Zhenxing Dai}
    \affiliation{
    Key Laboratory of Computational Physical Sciences (MOE), Fudan University, Shanghai 200433, China
    }
    \affiliation{Department of Physics, Fudan University, Shanghai 200433, China}
\author{Menglin Huang}
    \email{menglinhuang@fudan.edu.cn}
    \affiliation{
    Key Laboratory of Computational Physical Sciences (MOE), Fudan University, Shanghai 200433, China
    }
    \affiliation{College of Integrated Circuits and Micro-Nano Electronics, Fudan University, Shanghai 200433, China}
\author{Xin-Gao Gong}
    \affiliation{
    Key Laboratory of Computational Physical Sciences (MOE), Fudan University, Shanghai 200433, China
    }
    \affiliation{Department of Physics, Fudan University, Shanghai 200433, China}
\author{Shiyou Chen}
    \email{chensy@fudan.edu.cn}
    \affiliation{
    Key Laboratory of Computational Physical Sciences (MOE), Fudan University, Shanghai 200433, China
    }
    \affiliation{College of Integrated Circuits and Micro-Nano Electronics, Fudan University, Shanghai 200433, China}

\date{\today}

\begin{abstract}
Platinum (Pt) is widely used for carrier-lifetime control in silicon power devices, yet the microscopic nonradiative recombination mechanism of the substitutional platinum ($\text{Pt}_\text{Si}$) dopant remains debated. Using first-principles calculations combined with nonradiative multiphonon theory, we systematically investigate the electronic structures and carrier capture dynamics of $\text{Pt}_\text{Si}$. Our results show that both the donor ($+/0$) and acceptor ($0/-$) levels of $\text{Pt}_\text{Si}$ exhibit large capture cross sections for electron and hole carriers, thereby making $\text{Pt}_\text{Si}$ an effective recombination center. Notably, the calculated capture cross sections are sensitive to the symmetry-equivalent defect configurations with different Jahn-Teller distortions. By accounting for two different $D_{2d}$ configurations of neutral $\text{Pt}_\text{Si}$ during transitions properly, our calculated carrier capture cross sections align well with experimental values. This work provides a microscopic picture of the carrier capture processes induced by $\text{Pt}_\text{Si}$ and emphasizes the importance of symmetry-equivalent configurations in defect-assisted nonradiative recombination.

\end{abstract}

\maketitle

Nonradiative recombination at deep-level defects
strongly influences the performance of semiconductor devices \cite{Lutz2018,freysoldt2014first,park2018point}. In photovoltaic cells, defects that act as strong recombination centers limit carrier lifetimes and reduce the power conversion efficiency. In contrast, for power devices like insulated gate bipolar transistors (IGBTs), the precise introduction of recombination centers is essential to realize high switching speed and low turn‑off loss. Therefore, a quantitative understanding of the defect-assisted  nonradiative recombination is important for the optimization of device performance.

Platinum (Pt) impurities in silicon (Si) have long attracted considerable attention due to their application as carrier lifetime controllers. Compared to alternatives like gold, Pt offers superior high-temperature stability, better turn-on performance, and lower leakage currents~\cite{miller1976differences,baliga1977comparison}, which are fundamentally governed by the deep levels Pt introduces into the band gap. Specifically, donor levels within 0.26--0.36~eV above the valence band maximum (VBM) and acceptor levels within 0.19--0.28~eV below the conduction band minimum (CBM) have been widely observed in Pt-doped Si \cite{carchano1970electrical,conti1971electrical,pals1974properties,lisiak1975energy,miller1975use,pugnet1976energy,brotherton1979electrical,kwon1987properties,stoffler1986influence,gill1990role,zimmermann1991trivalent,deng1995platinum}. Although some studies suggested that the two levels may originate from distinct defects \cite{brotherton1979electrical,alves1986self}, these levels are widely attributed to substitutional Pt ($\text{Pt}_\text{Si}$) \cite{lisiak1975energy,pugnet1976energy,sachse1997electrical}. This assignment is further supported by the identification of a double donor level near the VBM, which is also assigned to the same platinum center \cite{zimmermann1991trivalent,sachse1997electrical}. 

While early studies \cite{miller1975use,pugnet1976energy,miller1976lifetime} proposed that the donor and acceptor levels of $\text{Pt}_\text{Si}$ serve as the dominant lifetime-controlling states in p-type and n-type Si, respectively, this view has been challenged by subsequent studies.
Lisiak et al. \cite{lisiak1975energy,lisiak1975platinum} argued that both levels of $\text{Pt}_\text{Si}$ would not be effective in recombination as they are far from the middle of the band gap. They proposed instead that a level at $E_v$ + 0.42~eV, possibly associated with a Pt-related complex, is more favorable to dominate the carrier lifetime. This interpretation was supported by Kumar et al. \cite{kumar1985dominant}, who observed a compensation effect induced by Pt doping and argued that an effective acceptor-type recombination center must be located near or below the intrinsic Fermi level of Si. Therefore, the $\text{Pt}_\text{Si}$ acceptor level, lying close to the CBM, was considered unlikely to be the dominant recombination level. 
Furthermore, Evwaraye et al. \cite{evwaraye1976electrical} proposed that another level at $E_c - 0.34$~eV in Pt-doped Si measured by deep-level transient spectroscopy (DLTS) is a good recombination center while the $\text{Pt}_\text{Si}$ acceptor level is not. These differing experimental interpretations therefore motivate a further examination of whether $\text{Pt}_\text{Si}$ can act as an effective recombination center.

Based on the Shockley-Read-Hall (SRH) theory \cite{shockley1952statistics}, assessing whether a defect acts as an effective recombination center requires determination of both majority and minority carrier capture cross sections. However, experimental data for $\text{Pt}_\text{Si}$, primarily from DLTS, provide only limited information on minority carrier capture and exhibit discrepancies of 2–3 orders of magnitude across different reports \cite{pals1974properties,lisiak1975energy,evwaraye1976electrical,brotherton1982measurement,zimmermann1991trivalent,Lutz2018,deng1995platinum}, hindering the definitive identification of recombination centers. Therefore, systematic theoretical evaluations of carrier capture cross sections are essential for clarifying the role of $\text{Pt}_\text{Si}$, yet such a study is still lacking.

In this study, we employ first-principles calculations to systematically investigate the electronic structure and nonradiative recombination mechanisms of $\text{Pt}_\text{Si}$ in Si. By calculating carrier capture cross sections based on the nonradiative multiphonon theory, we show that both donor (+/0) and acceptor (0/-) levels can facilitate rapid electron-hole recombination. The calculated capture process is sensitive to the choice of symmetry-equivalent defect configurations. By considering the $D_{2d}$ configurations of the neutral defect during transitions properly, we obtain capture cross sections in good agreement with experimental results. These results provide a microscopic picture to understand the nonradiative carrier capture processes induced by $\text{Pt}_\text{Si}$ and underscore the importance of symmetry-equivalent configurations when investigating defect-assisted recombination.

First-principles calculations are performed within density functional theory (DFT) using the Heyd-Scuseria-Ernzerhof (HSE) hybrid functional \cite{krukau2006influence} as implemented in the Vienna Ab Initio Simulation Package (VASP) \cite{kresse1996efficient}.  The projected augmented-wave (PAW) pseudopotentials \cite{kresse1999ultrasoft} are used with a plane-wave cutoff energy of 370~eV, and a single $\Gamma$-centered k-point mesh is used to sample the Brillouin zone. Spin polarization is included in all cases. The fraction of Hartree–Fock exchange is set to 0.25. The calculated band gap is 1.20~eV at the $\Gamma$ point, in good agreement with the experimental value of 1.17~eV \cite{bludau1974temperature}. During structural relaxation, all atoms in the supercell are fully relaxed until the force on each atom falls below 0.01~eV/\r{A}. 
Although Pt is a heavy $5d$ element, test calculations show that spin-orbit coupling (SOC) has negligible effects on the electronic structures. Therefore, SOC is neglected hereafter (see the Supporting Information, SI).

The formation energy $\Delta E_f(\alpha, q)$ for a defect $\alpha$ in charge state $q$ is determined as follows \cite{freysoldt2014first}:
\begin{equation}
\begin{aligned}
    \Delta E_f(\alpha, q) &= E(\alpha, q)-E(\text{host})+\sum_i n_i (E_i+\mu_i)\\ & + q(E_F+\varepsilon_\text{VBM})+E_\text{corr}
\end{aligned}
\end{equation}
where $E(\alpha, q)$ and $E(\text{host})$ are the total energies of the supercell with and without the defect, respectively. $\mu_i$ denotes the chemical potential of element $i$, referenced to the energy $E_i$ of the stable elemental phase, and $n_i$ is the number of atoms removed from ($n_i > 0$) or added to ($n_i < 0$) the supercell to form the defect. $E_F$ is the Fermi level referenced to the VBM of the host supercell. Finally, $E_\text{corr}$ is the finite-size correction for charged defects. We employ the correction scheme proposed by Freysoldt, Neugebauer, and Van de Walle (FNV) \cite{freysoldt2009fully}, using a static dielectric constant of 11.9.
Formation energies and transition levels are calculated within 512-atom supercells.
Considering the computational cost and the fact that nonradiative capture is mainly driven by local lattice distortion and electron–phonon coupling in the vicinity of the defect, carrier capture cross sections are calculated in a 216-atom supercell (size-convergence verified in the SI). We use the nonradiative multiphonon theory to calculate the capture cross section following the methodology described in Ref.~\onlinecite{zhou2025defect}, which explicitly considers the change of phonon modes due to lattice relaxation. 

\begin{figure}
\includegraphics[scale = 0.37]{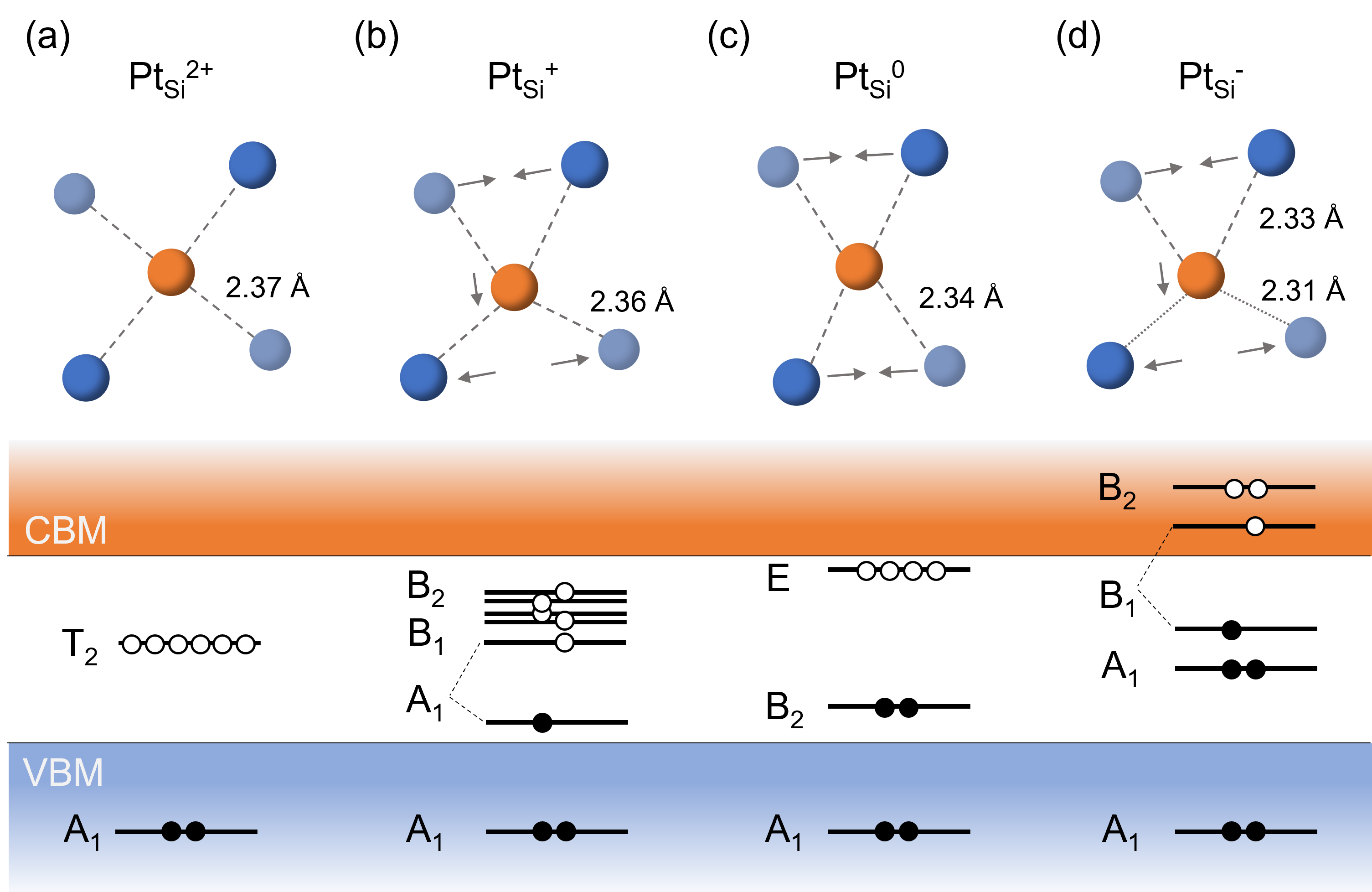}
\caption{\label{config} Equilibrium atomic structures of the platinum substitutional defect in the (a) double-positive, (b) single-positive, (c) neutral, and (d) single-negative charge states. Platinum atom is shown in orange and silicon atoms are in blue, with each unique Si-Pt bond length labeled. The corresponding electron energy level diagrams at the $\Gamma$ point with character symmetry labels are shown in the bottom panel.}
\end{figure}

We first examine the equilibrium atomic structures of the defects in different charge states. In a perfect Si crystal, a substitutional atom preserves $T_d$ symmetry. Figure~\ref{config} shows that this symmetry is maintained only for the double-positive state ($\text{Pt}_\text{Si}^{2+}$), where the Pt atom remains at the lattice site center and the four Pt–Si bond lengths increase from 2.35~\r{A} in bulk Si to 2.37~\r{A}. For the other three charge states, we observe significant structural distortions driven by the Jahn-Teller effect. The neutral state ($\text{Pt}_\text{Si}^0$) exhibits $D_{2d}$ symmetry, whereas the ground-state structures of single-positive ($\text{Pt}_\text{Si}^+$) and single-negative ($\text{Pt}_\text{Si}^-$) charge states both adopt $C_{2v}$ symmetry.

The observed structural distortions can be understood within the vacancy model proposed by Watkins \cite{watkins1983deep}. The defect levels in the band gap originate mainly from vacancy-like orbitals
of the neighboring Si atoms, while the Pt $5d$ states are located deep in the valence band and remain fully occupied.
The partially occupied $T_2$ levels in the gap are susceptible to Jahn–Teller distortions, driving the symmetry lowering observed for the $+$, $0$, and $-$ charge states.

In the double-positive case, the $T_2$ level is empty, preserving the $T_d$ symmetry and pushing the nearest-neighbor Si atoms slightly outward. In the single-positive state, the single occupied $T_2$ level is unstable, which in turn induces spontaneous Jahn-Teller distortion to $C_{2v}$ symmetry and yields a spin $S=1/2$. This distortion results in the elongation of one pair of opposite Si–Si distances and shortening of the other pair.

In the case of neutral state $\text{Pt}_\text{Si}^0$, two electrons occupy the $T_2$ level, which splits into a fully occupied $B$ state and unoccupied doubly degenerate $E$ states under $D_{2d}$ symmetry, giving a low-spin configuration $S=0$. For the single-negative state $\text{Pt}_\text{Si}^-$, the ground-state configuration has $S=1/2$ and $C_{2v}$ symmetry. Compared with $\text{Pt}_\text{Si}^+$, the Pt atom exhibits a larger displacement here, reaching 0.29~\r{A}. This structure was experimentally confirmed by electron paramagnetic resonance (EPR) in the 1960s \cite{woodbury1962spin}.

Fig.~\ref{form}(a) shows the calculated formation energies of $\text{Pt}_\text{Si}$ defect in Si. As can be seen, three defect transition levels are introduced in the band gap. The (2+/+) and (+/0) donor levels are located at 0.14~eV and 0.39~eV above the VBM, respectively. The (-/0) acceptor level is 0.27~eV below the CBM. The calculated transition levels above are in good agreement with experimental results, as shown in Fig.~\ref{form}(b). 

\begin{figure}
\includegraphics[scale = 0.43]{figs/formation_energy.png}
\caption{\label{form} (a) Calculated formation energies of the $\text{Pt}_\text{Si}$ defect as a function of the Fermi level ($E_F$). Solid lines represent the ground-state configuration for each charge state ($T_d^{2+}$, $C_{2v}^+$, $D_{2d}^0$, and $C_{2v}^-$), while dashed lines denote higher-energy metastable configurations. The vertical dashed arrows mark the thermodynamic transition levels. (b) Comparison between the calculated transition levels (this work) and experimental values \cite{pals1974properties,evwaraye1976electrical,brotherton1979electrical,brotherton1982measurement,deng1995platinum,Lutz2018,zimmermann1991trivalent,sachse1997electrical}. All energies are referenced to the VBM or CBM as indicated.}
\end{figure}

\begin{figure}
\includegraphics[scale = 0.52]{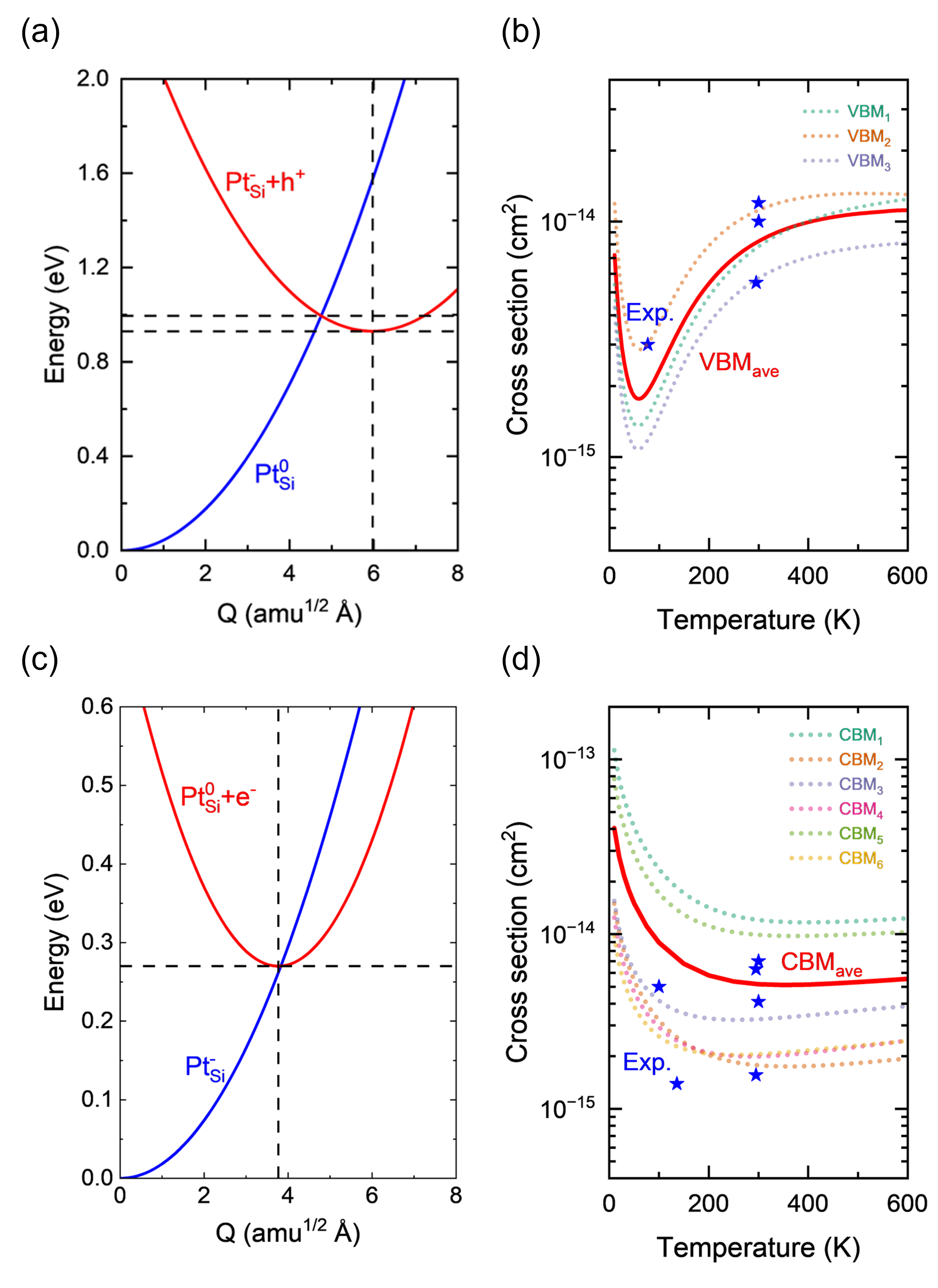}
\caption{\label{acceptor} (a) CC diagram for hole capture ($\text{Pt}_\text{Si}^- + \text{h}^+ \rightarrow \text{Pt}_\text{Si}^0$) to the $D_{2d}$ neutral state. (b) Calculated $\sigma_{pA}$ as a function of temperature, showing the average value (solid line) and contributions from three degenerate VBM states (dotted lines) compared with experimental data \cite{miller1976lifetime,Lutz2018, brotherton1982measurement, deng1995platinum} (stars). (c) CC diagram for electron capture ($\text{Pt}_\text{Si}^0 + \text{e}^- \rightarrow \text{Pt}_\text{Si}^-$) from the $D_{2d}^*$ neutral state. (d) Calculated $\sigma_{nA}$ compared with experimental data \cite{pals1974properties,brotherton1979electrical,brotherton1982measurement,deng1995platinum,evwaraye1976electrical,Lutz2018}. Solid and dotted
lines show the average and individual contributions from six CBM states.}
\end{figure}

To determine whether $\text{Pt}_\text{Si}$ defects can serve as effective recombination centers, we calculate the nonradiative capture cross sections of electrons and holes for different charge states of the defect.
We first analyze the nonradiative recombination process associated with the acceptor level (0/-). The configuration coordinate (CC) diagrams for these transitions are presented in Fig.~\ref{acceptor}, where the potential energy surfaces (PESs) of $\text{Pt}_\text{Si}^0$ and $\text{Pt}_\text{Si}^-$ are shown as functions of the generalized configuration coordinate $Q$ within the harmonic approximation.

In each diagram, $Q=0$ corresponds to the equilibrium configuration of the final state, and $\Delta Q$ denotes the displacement between the equilibrium configurations of the initial and final states.
The effective phonon energy of the final state is represented as the average of the phonon energies in different modes, weighted by the projection to lattice relaxation,
\begin{equation}
    \omega_i^2 = \sum_{\alpha,t,k}\omega_{ik}^2 (\mu_{\alpha t,0}\mu_{\alpha t,ik})^2
\end{equation}
where $\omega_{ik}$ is the energy of $k$-th phonon in the final state, and $\mu_{\alpha t,ik}$ is the corresponding phonon mode vector.
The effective phonon energy of the initial state based on the final-state phonon mode can be expressed as
\begin{equation}
    \begin{aligned}
        \omega_j^2 &= \sum_{\alpha,t,k'}\omega_{jk'}^2 (\mu_{\alpha t,0}\sum_{k}J_{ij,kk'}\mu_{\alpha t,ik})^2,\\
        J_{ij,kk'} &= \sum_{\alpha,t}\mu_{\alpha t,ik}\mu_{\alpha t,jk'}
    \end{aligned}
\end{equation}
where $J_{ij,kk'}$ is the Duschinsky matrix between final state phonon mode $k$ and initial state mode $k'$.

When $\text{Pt}_\text{Si}^-$ ($C_{2v}$) captures a hole from the VBM,  it transits into $\text{Pt}_\text{Si}^0$ ($D_{2d}$). In contrast, when $\text{Pt}_\text{Si}^0$ ($D_{2d}$) captures an electron from the CBM, the defect returns to the negatively charged state $\text{Pt}_\text{Si}^-$ ($C_{2v}$).
Within a semiclassical picture, carrier capture proceeds through the crossing point of the two PESs, where a thermal activation barrier may arise. 
As can be seen in the CC diagram for hole capture process shown in Fig.~\ref{acceptor}(a), the transition involves an energy barrier of 0.08~eV. Since the VBM of Si is triply degenerate, we calculate the hole capture cross sections ($\sigma_{pA}$) for each branch separately, together with their average value. As shown in Fig.~\ref{acceptor}(b), $\sigma_{pA}$ exhibits a non-monotonic temperature dependence. Below $100\text{ K}$, $\sigma_{pA}$ decreases with increasing temperature due to the reduction of the Sommerfeld factor \cite{passler1976relationships}, as the Coulomb attraction between $\text{Pt}_{\text{Si}}^-$ and the hole becomes less effective at higher carrier thermal velocities. As the temperature rises above $100\text{ K}$, $\sigma_{pA}$ increases as more phonons become available to promote the system toward the crossing point. This behavior is consistent with the temperature trend reported by Miller et al.~\cite{miller1976lifetime}.

We now turn to the electron capture process. 
Due to the indirect band gap of Si, the CBM consists of six equivalent valleys along the $\Delta$ directions. In the supercell calculation, these valleys are folded onto the $\Gamma$ point, giving six nearly degenerate conduction band states. We therefore calculated the electron capture cross sections ($\sigma_{nA}$) for each conduction band state and the averaged value.

Notably, we find that the electron capture process is sensitive to the Jahn-Teller distortion orientation of the initial and final states. Since both $\text{Pt}_\text{Si}^-$ ($C_{2v}$) and $\text{Pt}_\text{Si}^0$ ($D_{2d}$) can exhibit anisotropic distortions, multiple symmetry-equivalent orientations exist within the $C_{2v}$ and $D_{2d}$ distorted lattices. The hole capture transition [Fig.~\ref{acceptor}(a)] to the neutral $D_{2d}$ configuration exhibits a large lattice displacement ($\Delta Q \approx 6 ~\text{amu}^{1/2}$~\r{A}), involving a reorientation of the distortion direction relative to the $C_{2v}$ configuration (compared in Fig.~S2 of the SI). If the same $D_{2d}$ configuration is used for electron capture, the substantial structural reorganization would create a large barrier ($\approx 0.27$~eV) in the PESs, leading to a very small $\sigma_{nA}$ (see Fig.~S4 of the SI).
In contrast, if a symmetry-equivalent configuration, denoted as $D_{2d}^*$ [Fig.~\ref{acceptor}(c)], which shares a compatible distortion direction with the $\text{Pt}_\text{Si}^-$ state, is considered, the transition via this $D_{2d}^*$ configuration reduces the lattice displacement ($\Delta Q \approx 3.8~\text{amu}^{1/2}$~\r{A}) and leads to a nearly barrier-less transition, enabling a much more efficient capture pathway.

The calculated $\sigma_{nA}$ is shown in Fig.~\ref{acceptor}(d). The values have almost the same order of magnitude as that of $\sigma_{pA}$.
As the temperature rises from 0~K to 200~K, $\sigma_{nA}$ decreases from $4\times10^{-14} ~\text{cm}^2$ to $5 \times 10^{-15} ~\text{cm}^2$. Above 200~K, $\sigma_{nA}$ plateaus at the order of $10^{-15} ~\text{cm}^2$.
This negative temperature dependence is characteristic of barrier-less transitions, where thermal activation only yields limited increase in transition probability, causing the cross section to decline as the carrier thermal velocity increases ($\sigma \propto 1/v_{th}$).

It is important to note that the $D_{2d}$ and $D_{2d}^*$ configurations are thermodynamically degenerate, representing the same Jahn-Teller distortion along different axes. Climbing Image Nudged Elastic Band (CI-NEB) ~\cite{henkelman2000climbing} calculations reveal a low reorientation barrier of 0.12~eV between them (see Fig.~S3 of the SI), enabling rapid thermal reorientation on a sub-nanosecond timescale ($10^{10}$–$10^{11}$~Hz) at room temperature. Consequently, carrier capture kinetically funnels through the nearly barrier-less $D_{2d}^*$ pathway, which justifies evaluating the cross sections via this optimized configuration.

The large cross sections ($10^{-15} \text{--} 10^{-14} \text{ cm}^2$) for both electron and hole captures are in good agreement with experimental data, indicating that the acceptor level enables highly efficient recombination and further confirming that identifying the correct symmetry-equivalent configuration of Jahn-Teller distortion is essential for reproducing the microscopic capture dynamics of $\text{Pt}_\text{Si}$.

\begin{figure}
\includegraphics[scale = 0.52]{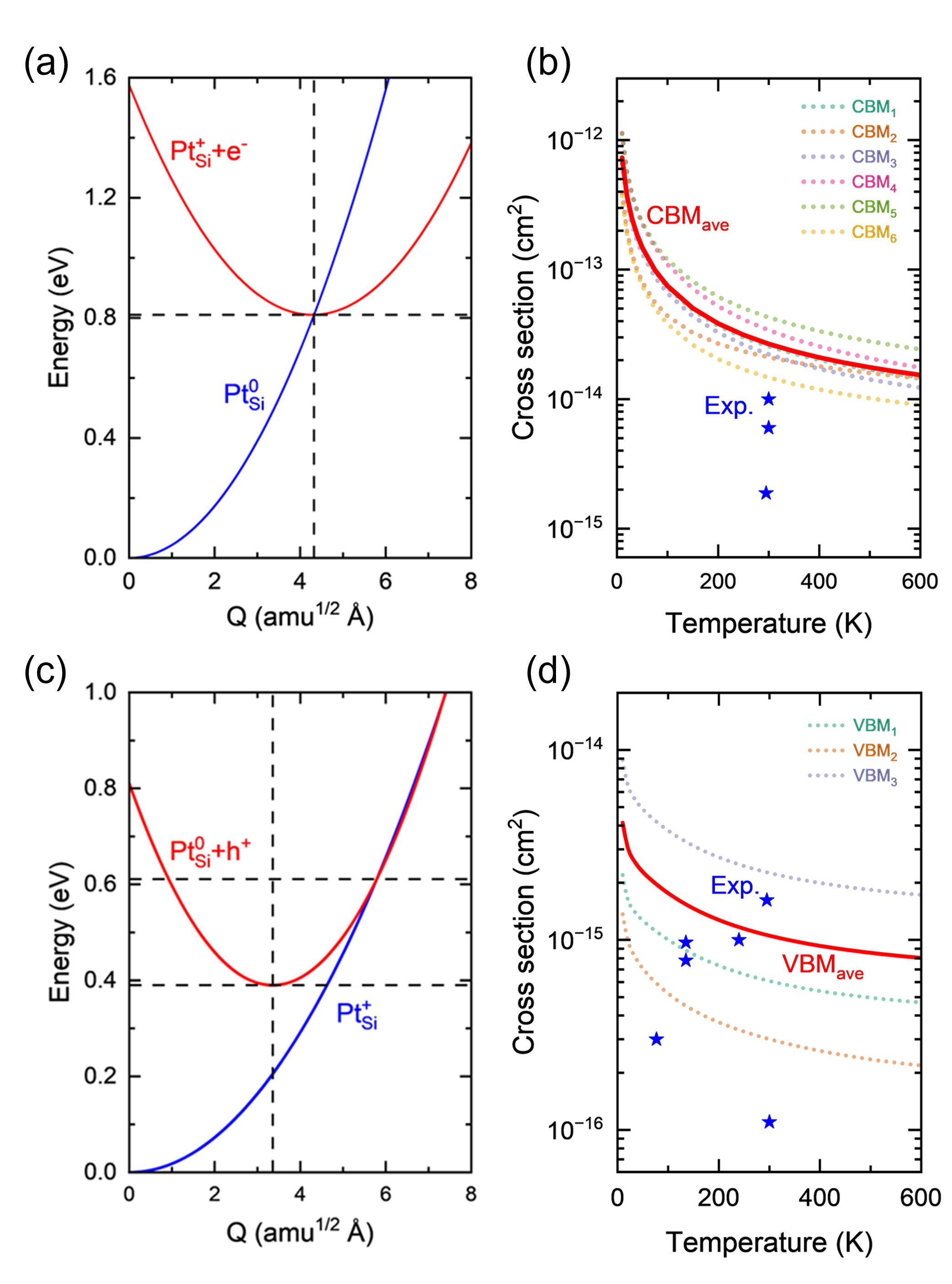}
\caption{\label{donor} (a) CC diagram for electron capture ($\text{Pt}_\text{Si}^+ + \text{e}^- \rightarrow \text{Pt}_\text{Si}^0$) to the $D_{2d}$ neutral state. (b) Calculated $\sigma_{nD}$ as a function of temperature, showing the average value (solid line) and contributions from six degenerate CBM states (dotted lines) compared with experimental data \cite{miller1976lifetime,Lutz2018,deng1995platinum} (stars). (c) CC diagram for hole capture ($\text{Pt}_\text{Si}^0 + \text{h}^+ \rightarrow \text{Pt}_\text{Si}^+$) from the $D_{2d}^*$ neutral state. (d) Calculated $\sigma_{pD}$ compared with experimental data \cite{pals1974properties,evwaraye1976electrical,brotherton1979electrical,brotherton1982measurement,deng1995platinum,Lutz2018} (stars). Solid and dotted lines show the average and individual contributions from three VBM states.}
\end{figure}

For the donor level (+/0), the nonradiative capture process also depends on the choice of symmetry-equivalent configuration.
As shown in Fig.~\ref{donor}(a), electron capture by $\text{Pt}_\text{Si}^+$ into the neutral $D_{2d}$ configuration proceeds via a nearly barrier-less transition, resulting in large electron capture cross sections ($\sigma_{nD} \ge 10^{-14} \text{ cm}^2$). Governed by Coulombic attraction, $\sigma_{nD}$ decreases monotonically with increasing temperature [Fig.~\ref{donor}(b)], consistent with the trend reported in Ref.~\onlinecite{miller1976lifetime}.

The subsequent hole capture process starting from the symmetry-equivalent $D_{2d}^*$ configuration of $\text{Pt}_\text{Si}^0$ involves a substantial activation barrier, as illustrated in Fig.~\ref{donor}(c).
The resulting capture cross section $\sigma_{pD}$ falls within a narrow range on the order of $10^{-15} \text{ cm}^2$ [Fig.~\ref{donor}(d)].
Despite the presence of the barrier, $\sigma_{pD}$ exhibits only a weak temperature dependence.
This behavior reflects the quantum-mechanical nature of multiphonon capture, where phonon-assisted tunneling effectively reduces the activation barrier inferred from the CC diagram \cite{alkauskas2014first,di1991advances}.
Overall, these large magnitudes of $\sigma_{nD}$ and $\sigma_{pD}$ (reaching $\ge 10^{-15} \text{ cm}^2$) confirm that the donor level, similar to the acceptor level, functions efficiently in capturing both types of carriers.

To explicitly connect our microscopic calculations to macroscopic performance, we evaluated the defect-assisted recombination efficiency of $\text{Pt}_\text{Si}$ at $300 \text{ K}$. Employing the recombination model based on the Sah-Shockley statistics~\cite{wang2025generalized}, we find the recombination coefficient $A$ reaches $\sim 10^7 \text{ s}^{-1}$ at a defect density of $10^{14} \text{ cm}^{-3}$, yielding a minority carrier lifetime below $100 \text{ ns}$. These results definitively validate that $\text{Pt}_\text{Si}$ acts as a highly efficient nonradiative recombination center. Further details are provided in the SI.

In summary, we have elucidated the microscopic nonradiative recombination dynamics of $\text{Pt}_\text{Si}$ in silicon using first-principles calculations combined with nonradiative multiphonon theory.
Our results show that $\text{Pt}_\text{Si}$ introduces $(2+/+)$, $(+/0)$, and $(0/-)$ levels within the band gap, consistent with experimental observations.
The calculated nonradiative capture cross sections indicate that both the donor $(+/0)$ and acceptor $(0/-)$ levels can efficiently capture electron and hole carriers, establishing $\text{Pt}_\text{Si}$ as an effective recombination center.
We further show that the carrier capture behavior is sensitive to the choice of symmetry-equivalent defect configurations with different Jahn-Teller distortions. These findings provide a microscopic understanding of carrier recombination at $\text{Pt}_\text{Si}$ and highlight the importance of selecting symmetry-equivalent configurations in quantitative descriptions of defect-assisted nonradiative capture process.

\section*{AUTHOR DECLARATIONS}
\subsection*{Conflict of Interest}
The authors have no conflicts to disclose.

\section*{Author Contributions}

\section*{Data Availability}
The data that support the findings of this study are available from
the corresponding author upon reasonable request.

\section*{References}

\nocite{*}
\bibliography{aipsamp}

\end{document}